\documentclass[manuscript]{aastex}

\shorttitle{LAT PSR J0357+3205 and its X-ray {\bf tail}}
\shortauthors{De Luca et al.}

\begin{document}


\title{Discovery
of a faint X-ray counterpart and of a parsec-long X-ray {\bf tail}
for the middle-aged, $\gamma$-ray only pulsar PSR J0357+3205}


\author{A.~De~Luca,\altaffilmark{1,2,3} M.~Marelli,\altaffilmark{2,4}
  R.~P.~Mignani,\altaffilmark{5,6}   P.~A.~Caraveo,\altaffilmark{2} 
W.~Hummel,\altaffilmark{7}
  S.~Collins,\altaffilmark{8} A.~Shearer,\altaffilmark{8}
 P.M.~Saz Parkinson,\altaffilmark{9} A. Belfiore,\altaffilmark{9,10}
%  P.~Ray (?),\altaffilmark{9} T.~Cheung (?),\altaffilmark{9} 
G.~F.~Bignami\altaffilmark{1,2}}

\altaffiltext{1}{IUSS - Istituto Universitario di Studi Superiori, viale Lungo Ticino Sforza, 56, 27100 Pavia, Italy}
\altaffiltext{2}{INAF - Istituto di Astrofisica Spaziale e Fisica Cosmica Milano, via E.~Bassini 15, 20133 Milano, Italy}
\altaffiltext{3}{INFN - Istituto Nazionale di Fisica Nucleare, sezione di Pavia, via A.~Bassi 6, 27100 Pavia, Italy}
\altaffiltext{4}{Universit\`a degli Studi dell'Insubria, Via Ravasi 2, 21100  Varese, Italy}
\altaffiltext{5}{Mullard Space Science Laboratory, University College London, Holmbury St. Mary, Dorking, Surrey, RH5 6NT, UK}
\altaffiltext{6}{Institute of Astronomy, University of Zielona G\'ora, Lubuska 2, 65-265 Zielona
G\'ora, Poland}
\altaffiltext{7}{European Southern Observatory, Karl Schwarzschild-Str. 2,
  D-85748, Garching, Germany}
\altaffiltext{8}{Centre for Astronomy, National University of Ireland,
  Newcastle Road, Galway, Ireland}
\altaffiltext{9}{Santa Cruz Institute for Particle Physics, Department of
  Physics, University of California at Santa Cruz, Santa Cruz, CA 95064, USA}
\altaffiltext{10}{Universit\`a degli Studi di Pavia, Dipartimento di Fisica
  Nucleare e Teorica, Via Bassi 6, 27100 Pavia, Italy}
%\altaffiltext{9}{Space Science Division, Naval Research Laboratory,
%  Washington, DC 20375, USA}
%\altaffiltext{8}{}
\email{deluca@iasf-milano.inaf.it}

%\author{S. Djorgovski\altaffilmark{1,2,3} and Ivan R. King\altaffilmark{1}}
%\affil{Astronomy Department, University of California,
%    Berkeley, CA 94720}

%\author{C. D. Biemesderfer\altaffilmark{4,5}}
%\affil{National Optical Astronomy Observatories, Tucson, AZ 85719}
%\email{aastex-help@aas.org}

%\and

%\author{R. J. Hanisch\altaffilmark{5}}
%\affil{Space Telescope Science Institute, Baltimore, MD 21218}


%\altaffiltext{1}{Visiting Astronomer, Cerro Tololo Inter-American Observatory.
%CTIO is operated by AURA, Inc.\ under contract to the National Science
%\altaffiltext{2}{Society of Fellows, Harvard University.}
%\altaffiltext{3}{present address: Center for Astrophysics,
%    60 Garden Street, Cambridge, MA 02138}
%\altaffiltext{4}{Visiting Programmer, Space Telescope Science Institute}
%\altaffiltext{5}{Patron, Alonso's Bar and Grill}

\begin{abstract}
The Large Area Telescope (LAT) onboard the {\em Fermi} satellite opened a new era
for pulsar astronomy, detecting $\gamma$-ray pulsations from more than 60
pulsars, $\sim40\%$ of which are not seen at radio wavelengths.  One of the
most interesting sources discovered by LAT is PSR J0357+3205, a radio-quiet, 
middle-aged ($\tau_C\sim0.5$ Myr)
pulsar standing out 
for its very low spin-down luminosity ($\dot{E}_{rot}\sim6\times10^{33}$ erg
s$^{-1}$), indeed the lowest among non-recycled $\gamma$-ray pulsars.
%as the non-recycled $\gamma$-ray pulsar with the lowest spin-down
%luminosity ($\dot{E}_{rot}\sim5\times10^{33}$ erg $^{-1}$) known so far.
A deep X-ray observation with Chandra (0.5-10 keV), coupled with sensitive optical/infrared 
ground-based images of the field, allowed us to identify PSR J0357+3205 
%in the 0.5-10 keV energy range
as a faint source with a soft spectrum, consistent with a purely non-thermal
emission (photon index $\Gamma=2.53\pm0.25$). The absorbing column
(N$_H=8\pm4\times10^{20}$ cm$^{-2}$) is consistent with a distance of a few
hundred parsecs.  
Moreover, the Chandra
data unveiled a huge (9 arcmin long) extended feature
apparently protruding from the pulsar. Its non-thermal X-ray spectrum 
points to synchrotron emission from energetic particles from the pulsar wind,
possibly similar to other elongated X-ray tails associated with rotation-powered pulsars and
explained as bow-shock pulsar wind nebulae (PWNe).
However, energetic arguments, as well as the peculiar morphology 
of the diffuse feature associated with PSR J0357+3205 
make the bow-shock PWN interpretation rather challenging.
\end{abstract}

\keywords{Stars: neutron --- Pulsars: general --- Pulsars: individual (PSR
  J0357+3205) --- X-rays: stars}

\section{Introduction}
The {\em Fermi} Large Area Telescope \citep[LAT, ][]{atwood09}, launched on June 11, 2008,
is revolutionizing our view of the high energy $\gamma-$ray sky,
thanks to its large collecting area and outstanding performance
at energies above 1 GeV.
One of the most exciting results obtained by 
{\em Fermi}-LAT has been
a factor of ten increase in the number of rotation
powered Pulsars (PSRs) identified as $\gamma$-ray sources. Starting from
a sample of 6 objects \citep[5 radio pulsars and Geminga, see e.g.][]{thompson08},
a legacy of the EGRET experiment onboard the Compton 
Gamma-ray Observatory, we have now
more than 60 identified pulsars \citep[][]{ray10,cara10},
%Ray \& Saz-Parkinson, 2010, arXiv:1007.2183; Caraveo, 2010, arXiv:1009.2421),
divided into three sub-families: classical pulsars (27 sources),
radio-quiet pulsars (22 sources) and millisecond pulsars (14 sources).
The existence of such population(s) of
gamma-ray pulsars  offers
a new view of the Galactic neutron stars and opens new avenues for neutron
star searches. While the wealth of detections confirms the importance
of the $\gamma$-ray channel
in the overall energy budget of rotation-powered pulsars, 
it points to emission
models in which the $\gamma-$ray production occurs in the outer
magnetosphere along open-field lines (outer gap / slot gap),
paving the way for understanding the 3-D structure and dynamics
of neutron star magnetospheres.

The most important objects to constrain
pulsar models are the ``extreme'' ones, accounting
for the tails of the population distribution in energetics,
age, magnetic field.
In this respect, PSR J0357+3205 is one of
the most interesting pulsars discovered by LAT. It is listed
in the catalog of the 205 brightest sources compiled after
3 months of sky scanning
\citep{abdo09a}, with a flux of $\sim1.1\times10^{-7}$
ph cm$^{-2}$ s$^{-1}$ above 100 MeV.
The source is located off the Galactic plane,
at a latitude $\sim-16^{\circ}$. A blind search allowed to
unambiguously detect the timing signature of a pulsar,
with P$\sim0.444$ s and $\dot{P}\sim1.3\times10^{-14}$ s s$^{-1}$
(Abdo et al. 2009b; see Ray et al. 2010 for the most updated timing parameters).
%PSR J0357+3205 is almost an order of magnitude less energetic 
%- and significantly older - than  Geminga,
%with an $\dot{E}_{rot}\, \sim \, 6\times10^{33}$
The characteristic age of PSR J0357+3205
($\tau_C=5.4\times10^5$ yr) is not outstanding 
among $\gamma$-ray pulsars. The ``Three Musketeers'' 
(Geminga, PSR B0656+14, PSR B1055-52, see e.g. De Luca et al. 2005),
have ages in the 115-550 kyr range and are prominent $\gamma$-ray sources
(Geminga and PSR B1055-52 are known to pulsate in $\gamma$-rays since
the EGRET era).
However, the spin-down luminosity of PSR J0357+3205
is as low as   $\dot{E}_{rot}\, = \, 5.8\times10^{33}$
erg s$^{-1}$, which is almost an order of magnitude lower than that 
of the Three Musketeers. Indeed, PSR J0357+3205
is the non-recycled $\gamma$-ray pulsar with the smallest rotational energy loss
detected so far.
This suggests PSR J0357+3205 to be rather close to us: by scaling
its $\gamma-$ray flux using the so-called $\gamma$-ray ``pseudo-distance'' relation
\citep[see e.g. ][]{sazparkinson10},
a distance of $\sim500$ pc is inferred.
%as distant as (or closer than) Geminga. 
PSR J0357+3205  shows
that even mature pulsars with a rather low spin-down luminosity
can sustain copious, energetic particle acceleration in their magnetosphere,
channelling a large fraction of their rotational energy loss in gamma rays. 
%As a matter of fact,
%existence of mature, low $\dot{E}_{rot}$ gamma-ray pulsars
%may only be explained within very recent models, e.g.
%the outer gap model by \citet{zhang04}.
Thus, it stands out as a powerful testbed for pulsar models.

In view of its plausible proximity, PSR J0357+3205 is a natural 
target for X-ray observations.
%The detection and identification of the X-ray counterpart of
%this peculiar source is a very challenging task. 
The lack of any plausible counterpart in
7 ks archival Swift/XRT data coupled with the quite large uncertainty in the position
of the $\gamma$-ray pulsar available in an earlier phase of the {\em Fermi} mission
%suggested us to use a deep multiwavelength (X-ray and optical) approach, 
\citep[$\sim5$ arcmin, ][]{abdo09b}
called for deep X-ray and optical observations
in order to identify the pulsar counterpart 
as an X-ray source with a high X-ray to optical flux ratio. This requires
(i) a deep X-ray observation with sharp angular resolution
to nail down the position of faint X-ray sources with sub-arcsec
accuracy and (ii) a sensitive multicolor optical coverage
of the X-ray sources detected inside the $\gamma-$ray error circle.
This second step is
crucial to reject unrelated field sources such as stars or
extragalactic objects.
To this aim, in 2009, we were granted of a joint
Chandra (80 ks) and NOAO program (4 hr in the V band and 3 hr in the Ks band 
at the Kitt Peak North Mayall 4m Telescope).
%We also collected optical images in the B, R and I bands at the
%... telescope thanks to the OPTICON Consortium \ref[REFERENCE][]{xx}.
We also made use of optical images in the B, R and I bands collected at the 2.5
m Isaac Newton Telescope (INT) at the La Palma Observatory (Canary Islands) in
2010 as a part of an International Time Programme aimed at a first follow-up of
{\em Fermi} $\gamma$-ray pulsars (Shearer et al., in preparation).

In this paper, we describe our observations, which yielded 
the X-ray counterpart of the pulsar 
as well as the detection of a huge extended feature
apparently linked to it.

\section{Observations}
\subsection{X-ray observations and data reduction}
Our Chandra observation of PSR J0357+3205 
was split between two consecutive
satellite revolutions. The first observation started on 25 October 2009 at
00:56 UT
and lasted  29.5 ks;
the second observation started on 26 october 2009 and lasted 47.1 ks. The two observations
are almost co-aligned, with very similar pointing directions and 
satellite roll angles. The target position
was placed on the back-illuminated ACIS S3 chip. The time resolution 
of the observation is 3.2 s. The VFAINT exposure mode was adopted.
We retrieved ``level 1'' data from the Chandra X-ray Center Science Archive
and we generated ``level 2'' event files using the Chandra Interactive
Analysis of Observations (CIAO v.4.2)
software\footnote{\url{http://cxc.harvard.edu/ciao/index.html}}. 
We also produced a combined event file using the {\em merge\_all} 
script\footnote{\url{http://cxc.harvard.edu/ciao/threads/combine/}}.

\subsection{Optical observations and data reduction}
Deep optical and near infrared images of the field of PSR J0357+3205 were collected 
at the 4m Mayall Telescope
at Kitt Peak North National Observatory as a part of our joint Chandra-NOAO program.
Optical observations in the V band (``V Harris'' filter, $\lambda=5375$ \AA,
$\Delta \lambda=$945.2 \AA) were performed using the large-field
($36'\times36'$) MOSAIC CCD Imager \citep{jacoby98} on 2009, November 10$^{th}$. 
Sky was mostly clear, with a few thin cirrus. Seeing was always better than $1.1''$.
We obtained a first set of 5 exposures of 10 min each and a second set of 18 exposures
of 12 min each, for a total integration time 
of 4 hr 26 min. 55\% of the observations were performed in dark conditions, 
45\% had partially ($\sim43\%$) illuminated Moon,
about 85$^{\circ}$ away from the target position. We used a standard 5-point dithering pattern. 
%Data reduction was performed using a suite of software
%tools\footnote{\url{http://www.noao.edu/noao/noaodeep/ReductionOpt/frames.html}}
%implemented within the IRAF software\footnote{\url{http://iraf.noao.edu/}}.
We performed standard data reduction (bias subtraction and flat fielding), CCD
mosaic, and image co-addition using the package {\tt mscred}
available in IRAF\footnote{\url{http://iraf.noao.edu/}}.

In our resulting co-added image, point sources have a full width at half
maximum of $\sim1.0''$ close to the expected target position. An astrometric solution
was derived using more than 1000 stars from the Guide Star Catalog 2
\citep[GSC2.3, ][]{lasker08} 
with a r.m.s. deviation of $\sim0.25''$ across the whole field of view. 
Following \citet{lattanzi97}, and taking into account the mean positional error 
in the GSC2 source coordinates as well as the uncertainty on the alignement
of GSC2 
with respect to the International Celestial Reference Frame \citep{lasker08},
%we end with an accuracy of $0.29''$ for the absolute astrometry of our image. 
our absolute astrometric accuracy is $0.29''$.
In view of the non-optimal sky conditions, 
photometric calibration of the image was performed using a set 
of more than 400 unsaturated sources, also listed in the
GSC2.3 catalogue, taking into account the transformation
from the photographic band to the Johnson band \citep{russell90} assuming a 
flat spectrum as a function of frequency. The r.m.s. of 
the fit is $\sim0.12$ mag.

Near Infrared observations were performed at Mayall on 2010,
February 2, using 
the Florida Multi-Object Imaging Near-infrared 
Grism Observational Spectrometer \citep[FLAMINGOS, ][]{elston03},
having a field of view of $10'\times10'$, using the Ks filter 
($\lambda=2.16\,\mu$m, $\Delta \lambda=0.31\,\mu$m).
%We collected ... 
%(ADD CONCISE DESCRIPTION OF DATA REDUCTION)
Sky conditions were not optimal, with passing thin to moderate cirrus clouds.
Seeing was good, always better than 0.9$''$.
To allow for subtraction of the variable IR sky background,
observations were split in 15 sequences (stacks) of short dithered exposures 
with integration time of 30 s.
Data reduction has been performed using 
MIDAS\footnote{\url{http://www.eso.org/sci/data-processing/software/esomidas/}}
and SciSoft/ECLIPSE\footnote{\url{http://www.eso.org/sci/software/eclipse/}} packages.
The near IR raw science images have
been linearized, dark subtracted and flat fielded.  The flat fields
have been generated from science frames via median stacking to ensure
that the flat field was stable over the two hours observing time.
Data consist of fifteen stacks, each one composed of 16
or 25 jittered images on a $4\times4$ or $5\times5$ grid.  
In a first step, the reduced raw frames of a
stack have been co-added using the SciSoft {\em jitter} command.  The sky was
subtracted as a moving average from the frames before the co-addition step.
A bad pixel map, derived from the flat field, was used for masking these
pixels.  For some of the stacks the jitter offset values required manual
adjustments in order to improve the image alignment.  As the ambient
conditions became less stable in the second half of the observing run
a brighter correlation star had to be used  further away
from the center of the field of view.
As a last step, the fifteen intermediate products have been co-added to
generate the final deep image. This image is composed of 254 raw frames,
and corresponds to a total integration time of 2h 7min.
An astrometric solution was computed ($\sim0.2''$ accuracy), based on a set of
stars from 
%the 2MASS 
the Two-Micron All-Sky Survey \citep[2MASS,][]{skrutskie06},
catalog. Photometric calibration, owing to poor
sky conditions, was performed on the image using a set of 35 stars also 
listed in the 2MASS catalog, with a r.m.s. of 0.12 mag.

%We also obtained optical observations 
%at the ... telescope within the frame of ... thanks to the OPTICON
%consortium \citep[REFERENCE][]{xx}. We collected ... hr in the ... filters.
%(DATA REDUCTION - TO BE DESCRIBED ELSEWHERE?)
Additional optical observations of PSR J0357+3205 in the B
($\lambda=4298$ \AA, $\Delta \lambda=$1065 \AA), R ($\lambda=$6380 \AA, $\Delta
\lambda=$1520 \AA),
and I ($\lambda=8063$ \AA,$\Delta \lambda=$1500 \AA) bands were
obtained in dark time with the Wide Field Camera (WFC) at the 2.5
m Isaac Newton Telescope (INT) at the La Palma Observatory (Canary Islands)
on the
nights of January 16-17 2010, with seeing in the $1.1''-1.3''$ range 
(see Shearer et al., in preparation), for a total
integration time of 6000 s in each band. The WFC is a mosaic of four thinned
2048$\times$2048 pixel CCDs, with a pixel size of 0\farcs33 and a full field of
view of $34\farcm2 \times 34\farcm2$. To compensate for the 1$'$ gaps
between the CCDs and for the fringing in the I band, observations were split in
sequence of 600 s exposures with a 5-point dithering pattern. Data reduction was
also performed with IRAF.
 Our astrometric solution was computed using 13 
USNOB stars\footnote{The stars used in the astrometric solution were 1220-0055300,
1220-0055302,1220-0055314, 1220-0055341, 1220-0055345, 1220-0055354,
1220-0055361, 1220-0055377, 1220-0055381, 1221-0061652, 1221-0061691,
1221-0061693, 1221-0061695.}
with $0.26''$ accuracy. For photometric calibration 45 USNO-B1 stars
\citep{monet03}
were used for I band images, 30 for B and 16 for R.
%Astrometric solution was computed with the GSC2
%(0\farcs25 accuracy) and photometric calibration was performed against Landolt
%stars (fields SA95, 98, 104).

\section{Results}
\subsection{The X-ray counterpart of PSR J0357+3205}
In order to identify the X-ray counterpart of the $\gamma-$ray pulsar, 
we  searched 
the most recent {\em Fermi}-LAT timing error circle 
for X-ray sources
showing 
the expected signature of isolated neutron stars, i.e.
a very high X-ray to optical flux ratio.

\begin{figure}
\plotone{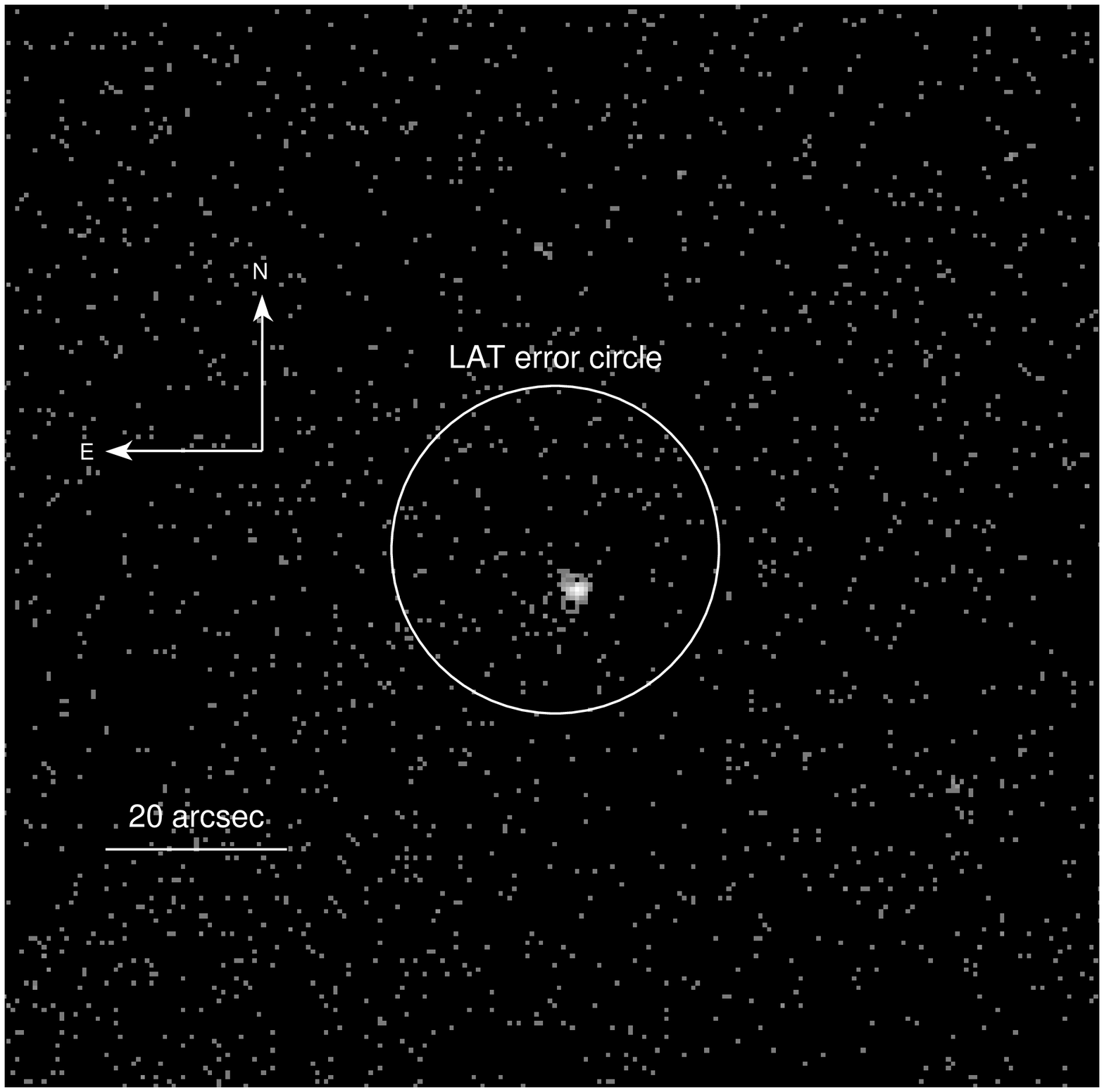}
\caption{Inner portion ($2'\times2'$) of the Chandra/ACIS 
image (0.5-6 keV) of the field of PSR J0357+3205. The {\em Fermi}-LAT 
timing error ellipse for the pulsar is superimposed.   
\label{chandrapsr}}
\end{figure}

\begin{figure}
\plotone{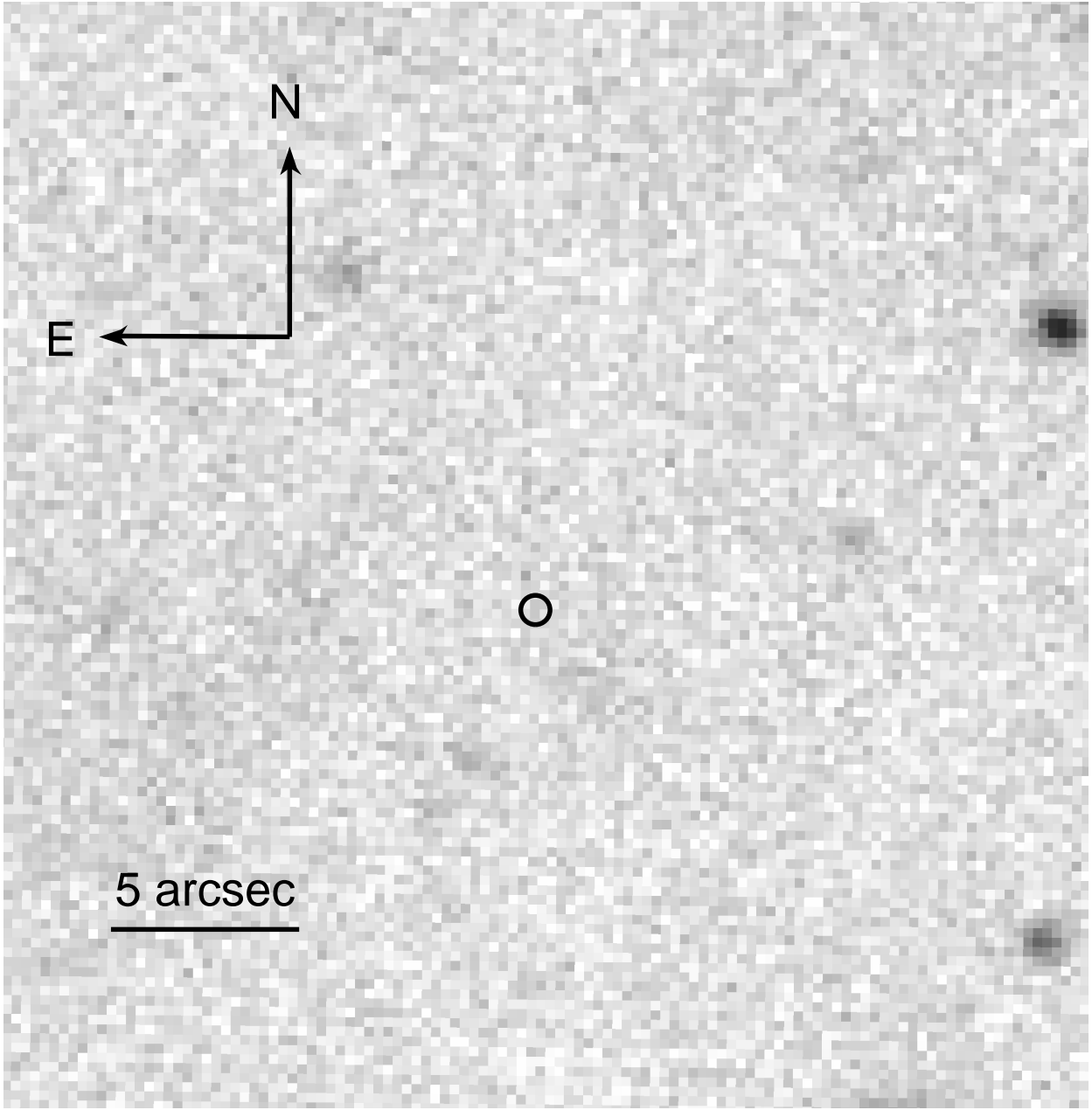}
\caption{
Inner region ($30''\times30''$) of the field as seen by the CCD Mosaic Imager
at the KPNO 4m telescope in the V band. Integration time is $\sim4.3$ hr. 
The circle marks the $1\sigma$ error circle ($0.4''$ radius) for the X-ray source
consistent with the position PSR J0357+3205. Positional error accounts for 
the uncertainty in the absolute astrometry of both X-ray and optical images.
No sources are seen at the position of the Chandra source (nor within
$\sim2.5''$ from it), down to V$>26.7$.  
\label{kpnopsr}}
\end{figure}

We generated an X-ray image in the 0.5-6 keV energy range 
using the ACIS original pixel size (0\farcs492).  
We ran a source detection using the 
{\it wavdetect}\footnote{\url{http://cxc.harvard.edu/ciao/threads/wavdetect/}} 
task, with wavelet
scales ranging from 1 to 16 pixels, spaced by a factor $\sqrt{2}$. A
detection threshold of $10^{-5}$ was selected in order not to miss
faint sources. 
The {\em Fermi}-LAT
timing error circle for PSR J0357+3205 is centered at R.A.=03:57:52.5,
Dec.=$+$32:05:25 and has a radius
of $18''$  \citep{ray10b}. Only one X-ray source,
positioned at  R.A.(J2000)= 03:57:52.32, Dec=+32:05:20.6,
is detected within such region, with a background-subtracted 
count rate of $(6.3\pm0.3)\times10^{-3}$ cts s$^{-1}$ in the 0.5-6 keV
energy range (see Fig.\ref{chandrapsr}).
%Its coordinates are %
%R.A.(J2000)= 03:57:52.32, Dec=+32:05:20.6.
In order to check the accuracy of the Chandra/ACIS absolute astrometry,
we cross-correlated positions 
of ACIS sources detected at $>4.5\sigma$ within 3 arcmin from the aimpoint 
with astrometric 
catalogues. We found two coincidences in the GSC2.3, 
with offsets of $0.15''-0.3''$. One of such two sources is also
listed in 
2MASS,
%the Two-Micron All-Sky Survey \citep[2MASS,][]{skrutskie06}, 
with a $0.15''$ 
offset with respect to the Chandra position. 
%As a further step, we cross-correlated the same X-ray sources with optical sources 
%of our large-field V-band image (linked to GSC2 within $0.25''$). 
%We found 9 matches, with a r.m.s. difference
%of $0.27''$ among X-ray and optical positions.
Although we could not derive an improved astrometric solution, 
%(any transformation turned out to be not statistically significant), 
such an exercise suggests that the Chandra 
astrometry is not affected by any systematics in our observations. 
Thus, we attach to the coordinates of our candidate counterpart
a nominal positional error of $0.25''$ (at $68\%$ confidence 
level\footnote{\url{http://cxc.harvard.edu/cal/ASPECT/celmon/}}). 
No coincident optical/infrared sources were found 
in our deep images
collected at Kitt Peak,
down to $5\sigma$ upper limits V$>26.7$, Ks$>19.9$ (the inner portion of the 
field, as seen in the V band, is shown in Fig.\ref{kpnopsr}). 
The INT telescope observation allows us to set 5 $\sigma$ upper limits 
of B$>25.86$, R$>25.75$ and I$>23.80$ (see Fig.~\ref{int}).
Assuming the best fit spectral model for the X-ray source (see below),
the corresponding X-ray to optical (V band) flux ratio is F$_X$/F$_V>520$,
while the X-ray to near infrared (Ks band) flux ratio is F$_X$/F$_{Ks}>30$.
%Lack of an infrared counterpart down to Ks$>...$ allows to exclude the 
%possibility of an extremely red extragalactic object as the 
Thus, positional coincidence coupled to very high X-ray to optical flux ratio prompt us to
conclude that our X-ray source is the 
counterpart of PSR J0357+3205.

To evaluate the source spectrum, we extracted photons within a $1.5''$ arcsec radius 
(561 counts in the 0.2-6 keV range, with 
a background contribution $<0.4\%$)
and we generated an ad-hoc
response matrix and effective area file using the CIAO script 
{\em  psextract}\footnote{\url{http://cxc.harvard.edu/ciao/threads/psextract/}}.
%The small number of counts from the pulsar , it is not possible
%to obtain a very detailed spectral characterization. 
We used the C-statistic
approach \citep[see e.g. ][]{humphrey09} implemented in 
XSPEC\footnote{\url{http://heasarc.gsfc.nasa.gov/docs/xanadu/xspec/XspecManual.pdf}} 
(requiring neither spectral grouping, nor background subtraction),
well suited to study sources with low photon statistics. Errors are at
$90\%$ confidence level for a single parameter.
The pulsar emission is well described
(the p-value, i.e. probability of obtaining the data if the model is correct, is 0.62) 
by a simple power law model, with a steep 
photon index ($\Gamma=2.53\pm0.25$),
absorbed by a hydrogen column density $N_H=(8\pm4)\times10^{20}$ cm$^{-2}$. A blackbody model
yields a poor fit (p-value $<0.00005$).
%($\chi^2_{\nu}=2.67$, 14 dofs). 
Assuming the best fit power law model, the 0.5-10 keV observed flux is 
$(3.9^{+0.7}_{-0.6})\times10^{-14}$ erg cm$^{-2}$ s$^{-1}$. The unabsorbed 0.5-10 keV flux 
is $4.7\times10^{-14}$ erg cm$^{-2}$ s$^{-1}$.

The limited statistics prevent us from constraining a more complex composite, thermal
plus non-thermal model, due to spectral parameter degeneracy
(e.g. $N_H$ vs. the normalization of the pulsar emission components).
To ease the problem, we can set an independent upper limit to the $N_H$.
The total Galactic absorption in the direction of the target is $(7-10)
\times 10^{20}$ cm$^{-2}$ \citep{dickey90,kalberla05}. 
Since such values, based on HI surveys, could differ significantly
with respect to the actual X-ray absorption,
we used our X-ray data to get an independent $N_H$ estimate.
%{\bf To get an estimate from our X-ray data}, we studied
%extra-galactic objects in the field. The 
Our brightest point source 
%in the field 
(source ``A'',
see fig.~\ref{trail})
is a {\it bona fide} AGN,
with a power law spectrum ($\Gamma=1.75\pm0.15$), a flux 
of $\sim1.1\times10^{-13}$ erg cm$^{-2}$ s$^{-1}$ and a F$_X$/F$_{opt}$ ratio
of $\sim11$. 
%is a {\it bona fide} AGN. 
Its absorbing column is
N$_H=(1.0\pm0.3)\times10^{21}$ cm$^{-2}$. Thus, we can assume conservatively
N$_H=1.3\times10^{21}$ cm$^{-2}$ as the maximum
possible value for the absorption towards the target. 
Such a constraint on N$_H$ allowed us to estimate upper limit temperatures
for any thermal emission from PSR J0357+3205
originating (i) from a hot polar cap  and (ii) from the whole 
neutron star surface. Assuming standard blackbody emission and
the standard polar cap radius 
%was assumed to have the radius of the
%footprints of last open field lines 
($r_{PC}=(2 \pi R^3 / c P)^{1/2}=320$ m), 
we obtain kT$<122$ eV (at 3$\sigma$ confidence level) 
for a 500 pc distance.
Similarly, for a NS radius of 13 km, we obtain 
kT$<$35 eV as a limit to the temperature of 
the whole surface of the star (at 3$\sigma$ confidence level).
Blackbody radii and temperature reported above are the values 
as seen by a distant observer.
%corresponding to 390 m as seen by a distant observer (for 
%$M_{NS}=1.4\,M_{\odot}$, r$_{NS}$=13 km), we obtain kT$<139$ eV
%(at 3$\sigma$ confidence level) for a 500 pc distance.

\subsection{The extended tail of X-ray emission}

Our Chandra data unveil the existence of a peculiar X-ray feature 
in the field of PSR J0357+3205. An extended structure of diffuse
X-ray emission, apparently protruding from the pulsar position,
is seen in the ACIS image, extending $>9$ arcmin in length 
and $\sim1.5$ arcmin in width (see Fig.~\ref{trail}). 
%Such a feature (the ``trail'' hereinafter) has 
A total of $1550\pm75$ background-subtracted counts
in the 0.5-6 keV band are collected from
such feature (the ``tail'', thereafter). 

We studied the morphology of the tail, 
extracting surface brightness profiles on different regions (see Fig.~\ref{regions}).
First, we searched for diffuse emission in the pulsar surroundings, 
by comparing the source intensity profile to the expected ACIS
Point Spread Function (PSF). Assuming the pulsar best fit spectral model,
we simulated a PSF using the
ChaRT\footnote{\url{http://cxc.harvard.edu/chart/threads/index.html}} 
and MARX\footnote{\url{http://cxc.harvard.edu/chart/threads/marx/} - We set
  the {\em DitherBlur} parameter to the value of $0.25''$ (smaller than
the default value of $0.35''$) in order to obtain a better reproduction of the
shape of the PSF in the inner core, as discussed by \citet{misanovic08}.}
packages. Results in the 0.5-6 keV energy range are shown in
Fig.~\ref{psf}, where the lack of any significant diffuse emission 
within 20$''$ of the pulsar position is apparent.
Then, we extracted exposure-corrected, background-subtracted 
surface brightness profiles on a larger angular scale.
Along the main (North-West to South-East) axis, 
the tail emerges from background $\sim20$ 
arcsec away from PSR J0357+3205, shows a broad
maximum after $\sim 4'$  
and then fades away at more than $9'$ from the pulsar.
(see Fig.~\ref{length}). 
A possible local minimum in the
surface brightness is also seen at $\sim2'$ from the pulsar position.
In the direction orthogonal to the main axis, 
the profile shows a sharper edge towards North-East
(rising to the maximum within 15$''$) and a shallower decay to the South-West
(fading to background in $\sim70''$), as shown in Fig.~\ref{width}. 
We also extracted energy-resolved images in
``soft'' (0.5-1.5 keV) and ``hard'' (1.5-6 keV) energy bands. However, no significant
differences in the brightness profiles are observed (see Fig.~\ref{length} and
Fig.~\ref{width}).

Spectral analysis of the tail emission is hampered by the low signal-to-noise
ratio. In the extraction region, background accounts for $\sim57\%$ of the
total counts in 0.5-6 keV. A background spectrum was extracted from a source-free
region north-east of the tail. Response and effective area files were
generated using the CIAO {\em specextract}
script\footnote{\url{http://cxc.harvard.edu/ciao/threads/specextract/}}.
The spectrum of the tail is described well ($\chi^2_{\nu}=1.0$, 70 dofs) by 
a non-thermal emission model (power law photon index $\Gamma=1.8\pm0.2$),
absorbed by a column $N_H=(2.0\pm0.7)\times10^{21}$
cm$^{-2}$. 
Confidence contours for $N_H$ and photon index of the diffuse 
feature, compared to the ones of the pulsar, are shown in Fig.~\ref{contours}.
%which is somewhat higher than the value computed
%for the pulsar counterpart.  However, we note that 
Fixing $N_H$ to
 $8\times10^{20}$ cm$^{-2}$ (best fit value for the pulsar counterpart)
yields a statistically
acceptable fit ($\chi^2_{\nu}=1.1$, 71 dofs), with a photon index 
$\Gamma=1.55\pm0.15$.
Adopting the latter model, 
the total observed flux of the tail
in the 0.5-10 keV energy range
is $(2.4\pm0.4)\times10^{-13}$ erg cm$^{-2}$ s$^{-1}$, corresponding to an
average surface brightness of $2.5\times10^{-14}$ erg cm$^{-2}$ s$^{-1}$
arcmin$^{-2}$. The unabsorbed flux in the same energy range
is $2.9\times10^{-13}$ erg cm$^{-2}$ s$^{-1}$.
We note that a thermal bremsstrahlung model also fits well the data
($\chi^2_{\nu}=1.0$, 70 dofs)
with an absorbing column $N_H=(1.4\pm0.7)\times10^{21}$ cm$^{-2}$, 
but requiring an unrealistic temperature, kT=5.4$\pm$1.7 keV.
%Using the non-thermal model with $N_H=8\times10^{20}$ cm$^{-2}$, 
%the total observed flux of the trail
%in the 0.5-10 keV energy range
%is $2.7\times10^{-13}$ erg cm$^{-2}$ s$^{-1}$, corresponding to an
%average surface brightness of $2.5\times10^{-14}$ erg cm$^{-2}$ s$^{-1}$
%arcmin$^{-2}$. The unabsorbed flux in the same energy range
%is $2.9\times10^{-13}$ erg cm$^{-2}$ s$^{-1}$.
%The maximun surface brightness is of ... erg cm$^{-2}$ s$^{-1}$ arcmin$^{-2}$.

%Spectral analysis of the trail points to a N$_H$ value somewhat larger 
%than the limits for the case of the pulsar. 
%A joint pulsar+trail spectral analysis setting the N$_H$ at such value
%yields statistically acceptable results, 
%with N$_H=(1.8\pm0.7)\times10^21$ cm$^{-2}$ ($\chi^2_{\nu}=0.78$, 45 d.o.f. -- see Table~\ref{jointfit}). 
%Even fixing the absorbing column, no useful 
%constraints could be obtained on the existence of  a thermal emission 
%component in the pulsar spectrum.
%Such a value is broadly consistent with 
%the Galactic column in this direction. 

Spatially-resolved spectroscopy was also performed, 
using two separate extraction
regions, both along the tail and across it, assuming
a power law model. 
However, no significant spectral differences were found,
which is consistent with the results of our energy-resolved imaging reported
above. 
For instance, dividing the tail in two sections, for $N_H=8\times10^{20}$ cm$^{-2}$, we found 
%a hint 
%of spectral steepening as a function of the distance from
%the pulsar along the main axis, with the 
photon indexes 
%rising from 
$\Gamma=1.45\pm0.15$ in the first half of the tail 
%to 
and $\Gamma=1.60\pm0.15$ in the second half.

%Contribution to the flux by point sources is expected to be very small.
No point sources are detected superimposed to the tail,
with the exception of two objects located close to the SE end 
(``source 1'' and ``source ``2'' in Fig.~\ref{trail}). 
X-ray spectroscopy points to non-thermal 
emission spectra for such sources. 
Both sources have very likely optical
counterparts in our ground-based images. 
The resulting X-ray to optical flux ratio
is F$_X$/F$_V\sim13$ and F$_X$/F$_V\sim4.5$
for source 1 and source 2, respectively. 
Such results allow us to conclude that they are unrelated extragalactic
sources. 
Our ground-based images do not show any bright optical
source possibly associated to the tail, nor hints of correlated, diffuse
emission. 
We also retrieved and analyzed public data at radio wavelengths
from the NRAO VLA Sky Survey 
\citep[NVSS,][]{condon98}. The images at 1.4 GHz do not show any counterpart
for the tail. 
We could set upper limits of 6.1 mJy to the tail radio emission 
over the whole extension of the X-ray feature (T. Cheung, private communication). 
The tail has also been detected by Suzaku
(Y. Kanai, private communication), in a 40 ks long observation,
although such data could not resolve its shape, nor
yield a better characterization of its spectrum.

\section{Discussion} 
\subsection{The X-ray counterpart of PSR J0357+3205}
Our multiwavelength campaign allowed us to identify the faint X-ray counterpart
of the $\gamma$-ray only pulsar PSR J0357+3205. 
Bright in $\gamma$-rays \citep{abdo09a, abdo09b}, with a $\gamma$-ray
to X-ray flux ratio of $F_{\gamma}/F_X\sim1,300$, PSR J0357+3205
is an unremarkable X-ray source. Although the small photon statistics
does not allow us to draw firm conclusions, the non-negligible interstellar absorption
points to a distance of a few hundreds parsecs for the source, in broad agreement
with the value of $\sim500$ pc estimated by scaling its $\gamma$-ray flux,
using the $\gamma$-ray pseudo-luminosity relation by \citet{sazparkinson10}.

The ACIS spectrum is consistent 
with a purely non-thermal origin of the X-ray emission.
The 0.5-10 keV luminosity (at 500 pc) is $L_X=1.4\times10^{30}$ erg s$^{-1}$, 
accounting 
for $\sim2.4\times10^{-4}$ of the pulsar rotational energy loss
$\dot{E}_{rot}$, in broad agreement with the dependence of the 
X-ray luminosity of rotation-powered pulsars on the spin-down luminosity 
%$\dot{E}_{rot}$
\citep{becker97,possenti02,kargaltsev08a}. The photon index 
is significantly steeper than the typical value of $\sim1.8$ 
observed for middle-aged pulsars \citep{deluca05}.

%It was not possible to 
No thermal emission from the neutron star surface was detected. The
$3\sigma$ upper limit to the bolometric luminosity is
$\sim5\times10^{31}$ erg $s^{-1}$. Such limit can
be compared to
the bolometric luminosity of the well studied surface thermal
emission of the Three Musketeers, which have a characteristic age similar to
that of our
target. The upper limit to the thermal emission from PSR J0357+3205 
is a factor of 10 lower than the bolometric luminosity
of PSR B0656+14 and 
PSR B1055-52\footnote{The revision of the distance to PSR B1055-52 suggested
  by \citet{mignani10} would translate to a factor $\sim4$ smaller luminosity.}
\citep{deluca05}, but it is
comparable to the luminosity of Geminga \citep{caraveo04}. 
Although PSR J0357+3205 turns out to be
the coldest neutron star in its age range (0.1-1 Myr), the upper limit
to its thermal emission  
is 
%not surprising for a $\sim$million-year old pulsar, being 
consistent with the expectations of several
cooling models \citep[see, e.g.,][]{tsuruta09,page09}. 
On the other hand, the apparent lack of
emission from the polar caps is also interesting, 
since PSR J0357+3205 is a bright $\gamma$-ray
pulsar, channelling about 40\% of its spin-down
luminosity in $\gamma$-rays of magnetospheric origin and thus
polar cap re-heating by ``return currents'' in the magnetosphere 
would be expected. 
Our limit to the temperature
of a hot polar cap points to a bolometric luminosity
$L_{PC}<5\times10^{30}$ erg s$^{-1}$, 
a factor $>5$ lower than the polar cap luminosity 
for PSR B0656+14 and PSR B1055-52 \citep{deluca05}, but
a factor $\sim10$ larger than the polar cap luminosity of Geminga \citep{caraveo04}.
The upper limit is a factor of a few lower than the luminosity 
expected by heating models based on return currents
of e$^{+}$/e$^{-}$ generated above the polar caps
by curvature radiation photons,
but it is consistent with expectations for polar cap heating
due to bombardment by particles created only by inverse Compton
scattered photons \citep{harding01,harding02}. 
PSR J0357+3205 is close to the death line for production of
e$^{+}$/e$^{-}$ by curvature radiation photons \citep{harding02},
which could explain the reduced polar cap heating. As a further 
possibility, the system's viewing geometry could play some role,
as in the case of Geminga, where the emitting area  and luminosity
of the thermally emitting polar cap suggested an almost aligned
rotator, seen at high inclination angle \citep{caraveo04,deluca05}.

When compared to other well-known
middle-aged rotation-powered pulsars,
the X-ray spectrum of PSR J0357+3205 is
remarkably different. 
Indeed, it is reminiscent of a number of {\em older} ($\tau_C \sim 10^{6}-10^{7}$ yr) pulsars, such as, e.g.,
PSR B1929+10 \citep{becker06}, B1133+16 \citep{kargaltsev06}, B0943+10
\citep{zhang05}, B0628+28 \citep{becker05}. 
A non thermal origin for the bulk of the X-ray emission from such
pulsars was proposed by \citep{becker04,becker06}, although such a picture
was questioned, e.g., by \citet{zavlin04} and \citet{misanovic08}, 
who preferred a composite, thermal plus
non-thermal spectral model.

\subsection{The X-ray tail}
The morphology of the tail, apparently protruding from
PSR J0357+3205 and smoothly connected to the pulsar counterpart
%coupled to the lack of any other obvious explanation of its nature,
strongly argues for a physical association of the two systems.
This is also supported by the lack of any other source 
possibly related to the extended feature. 
 Sources ``1'' and ``2'' are extragalactic objects. An interpretation 
of the feature as an AGN jet, associated e.g. to Source 2, can be safely
discarded, owing to
the lack of radio emission for both the point source
and the diffuse feature, at variance with all known AGN jets \citep{harris06}.
Furthermore, the angular extent of the feature would imply an unrealistic 
physical size, unless the source is quite local (a huge 200 kpc-long 
jet would 
imply an angular scale distance 
of order 80 Mpc, assuming standard cosmological parameters), which would call for a rich
multiwavelength phenomenology (the host galaxy itself -- with an angular 
scale well in excess of 1 arcmin -- should be clearly
resolved in our ground-based optical images).

Assuming an association of the feaure  to PSR J0357+3205, the observed extension of the tail, at 
a distance of 500 pc, would correspond to a physical length of $\sim1.3$ pc
(assuming no inclination with respect to the plane of the sky).

A few elongated ``tails'' of X-ray emission associated to  rotation-powered
pulsars have been discovered by
\citet{gaensler04,mcgowan06,becker06,kargaltsev08b}.
Such features are interpreted within
the framework of bow-shock, ram-pressure dominated, pulsar wind nebulae \citep[see ][for a
review]{gaensler06}.
If the pulsar moves supersonically, shocked pulsar
wind is expected to flow in an
elongated region downstream of the termination shock (basically, the cavity in the interstellar medium
created by the moving neutron star and its wind),
confined by ram pressure. X-ray emission is due to synchrotron emission
from the wind particles accelerated at the termination shock, which 
%The termination shock, surrounding the pulsar on a smaller angular scale, 
is typically
seen (if angular resolution permits) as the brightest portion of the extended
structure \citep[see e.g. ][]{kargaltsev08b}, 
%with a ``bullet'' shape,
as expected from MHD simulations \citep{bucciantini02,vanderswaluw03,bucciantini05}.

Although for our radio-quiet pulsar we have no information about the proper
motion, the bow-shock PWN scenario would seem the most natural
explanation. Of course, such a picture would suggest for PSR
J0357+3205  a large space velocity aligned with the tail, in the direction opposite to the tail extension. 
If this is the case, the pulsar would be moving almost 
parallel to the plane of the Galaxy, which would suggest 
that it was born out of the Galactic plane, at an height 
of order 140$d_{500}$ pc (where $d_{500}$ is the distance to the pulsar 
in units of 500 pc), possibly from a
``runaway'' high mass star \citep{mason98}.

The luminosity of the tail in the 0.5-10 keV energy range
(assuming d=500 pc) is $8.8\times10^{30}$ erg s$^{-1}$, corresponding to 
a fraction $1.5\times10^{-3}$ of the pulsar spin-down luminosity. Indeed, such value
%lies precisely in the middle of the range $10^{-2}-10^{-4}$ 
is fully compatible with that measured for other pulsars, which channel
into their tails $10^{-2}-10^{-4}$ of their rotational energy loss.
%for other elongated features interpreted as ram pressure dominated PWNe.
Synchrotron cooling of the particles  
injected at the termination shock induces
a significant softening of the emission
spectrum as a function of the distance from the pulsar in Bow-shock PWNe. 
For the tail 
of PSR J0357+3205 we do not have firm evidence for such a spectral variation.
%although our data allow us to see hints of the expected steepening.

However, explaining the tail 
of PSR J0357+3205 within the bow-shock PWN frame is not straightforward.
A first difficulty arises from energetic requirements for 
the emitting particles --
indeed, the hypothesis that the observed X-rays
from the tail are due to synchrotron emission
is somewhat challenging for a pulsar with
such a low $\dot{E}_{rot}$. As
in the case of PSR B1929+10, discussed by \citet{becker06} and
\citet{dejager08}, the problem lies with the maximum 
energy of the particles injected in the PWN.
Particle acceleration mechanisms in PWNe are not yet fully
understood. The maximum energy to which electrons can be
accelerated (via acceleration of the pulsar wind 
and then re-acceleration at the termination shock)
is expected to be a fraction of the polar cap potential
($\sim0.1$ for the Crab, see de Jager et al.1996; see also de Jager \&
 Djannati-Ata\"i 2008, Bandiera 2008).
According to \citet{goldreich69}, 
the maximum potential drop between the pole and the light cylinder 
(in an aligned pulsar) is $\Delta \Phi = (3\dot{E}_{rot}/2c)^{1/2}$. 
For PSR J0357+3205, this would correspond to electron acceleration 
in the pulsar magnetosphere up to a maximum Lorentz factor
$\gamma_{max}\sim10^8$, which can be considered as an upper limit
for the electrons injected in the PWN. 
%Particle acceleration mechanisms in PWNe are not yet fully
%understood, and the maximum Lorentz factors fo leptons injected in the 
%PWN is generally expected to be smaller than $\gamma_{max}$ (ref).
The characteristic energy of synchrotron
photons is $\sim0.5 B_{-5} \gamma_8^2$ keV,
where $B_{-5}$ is the ambient magnetic field in units of 10 $\mu G$ and 
$\gamma_8$ is the Lorentz factor of the radiating electrons
in units of $10^8$. It is clear that, in order to produce bright emission
at few keV, the typical Lorentz factor of the electrons 
in the tail has to
be of the same order of $\gamma_{max}$, 
implying the presence of e$^+$/e$^-$ accelerated 
at the highest possible energy, as well as
%i.e. of maximally accelerated electrons from the pulsar magnetosphere. But
%this is not enough: 
an ambient magnetic field 
as high as $\sim50$ $\mu G$. 
 If this is the case, it is possible to estimate 
the synchrotron cooling time of the emitting electrons
as $\tau_{sync}\,\sim100\,(B/50 \mu G)^{-3/2}\,(E/1\,keV)^{-1/2}$ yr.
Coupling such value with the estimated physical length of the feature
yields an estimate of the bulk flow speed of the emitting particles
of $\sim15,000$ km s$^{-1}$, assuming no inclination w.r.t. 
the plane of the sky. Such a value is consistent with results 
for other bow-shock PWNe \citep{kargaltsev08b}.

A second difficulty for the bow-shock interpretation arises owing to the 
lack of diffuse emission surrounding the pulsar.
Bright emission from
the wind termination shock should be seen there as 
%the portion of the diffuse feature with
the maximun surface brightness portion of the diffuse feature,
as observed in all other known cases 
\citep[see e.g.][]{gaensler04,mcgowan06,kargaltsev08b}.
As a possible way out, we evaluate under what conditions the termination shock could be unresolved by Chandra.
Assuming standard relations \citep{gaensler06}, the distance between the 
pulsar and the head of the termination shock is expected to be
$r_S=(\dot{E}_{rot}/4 \pi c
\rho_{ISM} v^2_{PSR})^{1/2}$, where $\rho_{ISM}$ is the ambient density
and $ v_{PSR}$ is the pulsar space velocity.
For PSR J0357+3205,
$r_S\sim10^{16}v_{PSR,100}^{-1}n_{ISM,1}^{-1/2}$ cm, 
where $v_{PSR,100}$ is the 
pulsar space velocity in units of 100 km s$^{-1}$ and $n_{ISM,1}$ is the 
ambient number density in units of 1/cm$^3$. At a distance of 500 pc, 
this corresponds to $\sim1.3''v_{PSR,100}^{-1}n_{ISM,1}^{-1/2}$.
The surface of the 
termination shock (in the hypothesis of an isotropic pulsar wind) 
should assume an elongated
shape, extending $\sim6 r_S$ ($\sim8''v_{PSR,100}^{-1}n_{ISM,1}^{-1/2}$ at 500
pc) behind the pulsar. The termination shock could
hide within the point spread function of the pulsar if $6r_s<0.5''$, which would require
an unrealistically large ambient number density (of order several hundreds per
cm$^3$), and/or a pulsar space velocity of at least 1000 km
s$^{-1}$. Anisotropies in the pulsar wind could also play some role.
Such a picture would suggest that a significant fraction
of the flux of the point source is due to emission from the wind 
termination shock.

A further problem with the bow-shock picture is related to the brightness
profile of the tail, which  is remarkably different from what is observed for all
other diffuse structures interpreted as ram-pressure dominated PWN. Figures 3
and 4 clearly show that
the surface brightness grows as a function of angular distance from PSR J0357+3205 
and reaches a broad maximum $\sim4'$ away from the pulsar, while 
%in all other cases the peak of 
all the elongated structures imaged so far have their peak 
close to their parent pulsar position
(althougn localized, bright ``blobs''
along the tails have been observed, see e.g. Kargaltsev \& Pavlov 2008, and 
interpreted as due to kink instabilities in the particle flow). 
Invoking geometric effects, such as bending of the tail along the line of sight,
producing limb brightening (higher column density of emitting particles) 
would require ad-hoc assumptions for the tail 3-D structure. Since no
plausible explanations for the origin of such bending are apparent, we discard
such a possibility. Lack of any significant spectral evolution along the tail 
ultimately prevents us from drawing conclusions on the physical nature of its peculiar profile.

The ``asymmetric'' brightness profile of the tail 
in the direction perpendicular
to its main axis (with its sharp north-eastern edge and its shallower
decay towards South-West)
is also remarkably different from what is observed for any
other diffuse structure interpreted as ram-pressure dominated PWN, but
%the brightness profile of the tail along its minor axis
is reminiscent of the case of the peculiar diffuse X-ray feature
associated to PSR B2224+65, powering the ``guitar'' nebula
\citep{hui07,johnson10}. 
The extended feature seen there is
remarkably misaligned (by $118^{\circ}$) wrt. the direction of the pulsar proper motion
\citep{hui07} and therefore it has been interpreted in a different frame, either as a 
``magnetically-confined'' PWN \citet{bandiera08}, or as a jet
 from the pulsar \citep{johnson10}. Both pictures naturally predict  
the feature to be brighter in the leading edge
(the one in the direction
of the proper motion), where ``fresh'' electrons are injected. 
The profile in the trailing edge is expected to fade smoothly,
dominated by cooling of the electrons deposited by the moving source
(i.e. the feature is a ``synchrotron wake'' along its minor axis). 
Thus, if this is the case, 
the proper motion of the pulsar 
should not be aligned with the tail main axis
and the tail itself should display a proper motion.
In view of the lack of information about the 
pulsar proper motion, it is premature to discuss 
such scenarii any more. We note, however, that
%that both pictures would suffer of important difficulties:
%the lack of any ``counter''-feature on the opposite side of the pulsar (expected 
%small (or no) bending
%Energetic electrons accelerated at 
%the pulsar wind termination shock leak out of the bow-shock and
%diffuse into a large-scale, plane-parallel magnetic field, emitting
%synchrotron radiation. The brighter edge is the one towards the direction
%of the proper motion, where ``fresh
%Moreover, 
 both pictures are not free from 
difficulties. For instance, the jet explanation cannot 
easily explain the lack of any appreciable bending of 
the structure due to ram pressure from the ISM. On the other  
hand, the magnetically confined PWN would require a very 
intense (50 $\mu$G), ordered ambient magnetic field 
\citep[see ][for further details on such pictures for the case of PSR B2224+65]{bandiera08,johnson10}.
Moreover, the broad maximum at a large distance from the pulsar 
would not be accounted for easily in these pictures (which, similarly to the
Bow-shock picture, predict
a brightness peak close to the pulsar). 

\section{Conclusions}
We have detected the faint X-ray counterpart of the middle-aged, $\gamma$-ray only, pulsar
PSR J0357+3205, together with an associated, elongated feature of diffuse
X-ray emission. The pulsar emission is consistent with a purely
magnetospheric, non-thermal origin. Future deep X-ray observations will allow
to better constrain the interstellar absorption (consistent with a distance 
of a few hundred parsecs) and possibly to detect pulsations. As for the case of the
$\gamma$-ray only pulsar in the CTA-1 supernova remnant \citep{caraveo10b},
this could unveil the presence of thermal emission from rotating hot spots,
possibly associated to polar cap reheating by magnetospheric currents.
The diffuse feature is $\sim9$ arcmin long (to our knowledge, 
considering its angular extension, this
is the largest ``tail'' of X-ray emission so far associated to 
a rotation-powered pulsar) and displays a hard, non-thermal spectrum. The nature 
of such feature cannot be firmly established. A crucial piece of information
could come from  the pulsar proper motion. In this respect, if 
the lack of a discernible pulsar wind termination shock is indeed due to 
a very high pulsar velocity ($\sim1000$ km s$^{-1}$), at a distance of 500 pc
this would translate to a proper motion of
$\sim0.42''$ yr$^{-1}$, a value which is within the reach of Chandra,
even with a short time baseline ($\sim2$ yr). We note that precise timing of LAT 
photons is not expected to be sensitive to the proper motion of PSR J0357+3205 
(Ray et al. 2010 estimated that timing based on 5 years of LAT data
will yield a positional accuracy of $\sim2''$). 
%which could be measured with future Chandra
%observations on a few year time span. 
A proper motion aligned with the tail
main axis would point to a bow-shock PWN interpretation, which will have, in
any case, to
face difficulties related to the energetics of the emitting particles as 
well as to the peculiar brightness profile.  
%Evidence for such a large angular displacement could 
%even come from timing of $\gamma$-ray photons collected by Fermi-LAT in a few years.
Conversely, a proper motion misaligned with respect to the tail axis
would point to a ``Guitar''-like system, to be interpreted as a 
magnetically confined PWN or as a pulsar jet. In such a case, 
proper motion of the tail itself could be detected. 
A further check on the tail nature could come from deep  X-ray observations,
which could allow to detect spectral steepening in its emission,
possibly shedding light on the geometry of the injection of particles
in the nebula and pointing either to the bow-shock scenario, or 
to the Guitar-like picture.
A long XMM observation, recently granted,  will clarify the spectral
behaviour 
of the pulsar as well as of its record long tail.

\acknowledgments
This work was partially supported by ASI-INAF contracts I/009/10/0
and I/047/08/0.
Based in part on observations taken as part of the CCI
International Time Programme 2009.
We warmly thank Buell Jannuzi and Heidi Schweiker for collecting the optical
images with the MOSAIC instrument, and Dick Joyce for performing
our infrared observations with the FLAMINGOS detector at the Kitt Peak North
Mayall Telescope.
We thank Teddy Cheung for calculating upper limits on the radio emission
of the tail using NVSS data and Yoshikazu Kanai for sharing Suzaku 
results included in his PH.D. Thesis. We also thank Paul S. Ray for useful discussions.

{\it Facilities:} \facility{CXO (ACIS)} \facility{Mayall (MOSAIC,FLAMINGOS)}.

\clearpage

\clearpage

\clearpage

\begin{figure}
\plotone{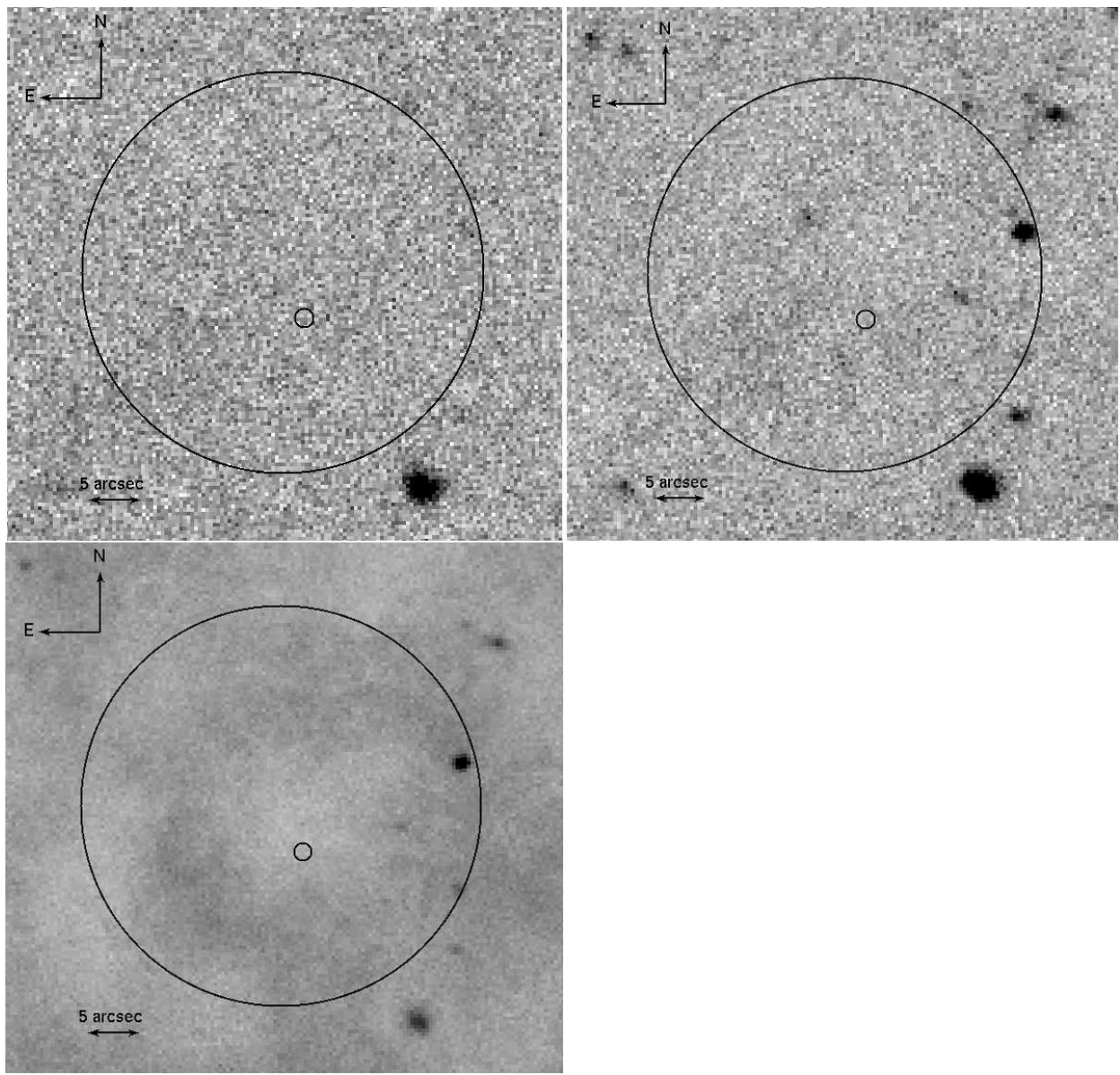}
\caption{Inner region ($40''\times40''$) of the field as seen by the WFC
on the INT
2.5m telescope in the B,R (top) and I (bottom) bands. Integration time is 6000
seconds in all cases. The smaller circle (0.8$''$ radius) marks the
$2\sigma$ error circle of the X-ray source consistent with the
position of PSR J0357+32. The larger circle shows the Fermi LAT 18$''$
error circle.
The positional error accounts for the uncertainty in the absolute
astrometry of both X-ray and
optical images. No sources are seen at the position of the Chandra
source (nor within $\sim2.5''$
from it), down to a 5 $\sigma$ limit of B$>$25.86, R$>$25.75 and I$>$23.80.
\label{int}}
\end{figure}

\clearpage

\begin{figure}
%\epsscale{.80}
\plotone{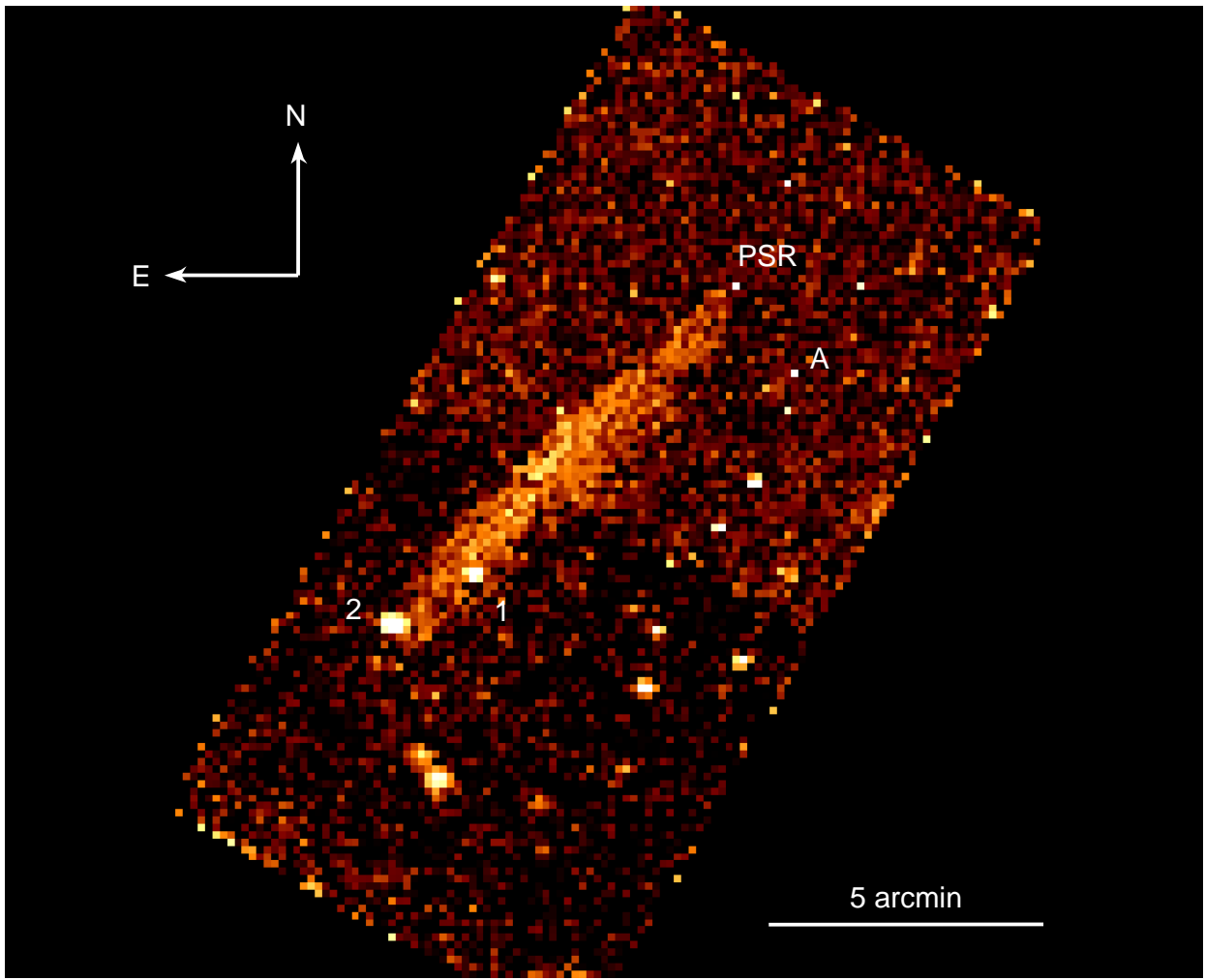}
\caption{Exposure corrected Chandra/ACIS image of the field of PSR J0357+3205
in the 0.5-6 keV energy range. The image has been rebinned to a pixel scale
of $8''$. No smoothing has been applied. A large tail of diffuse X-ray
emission is apparent, with a length (North-West to South-East) of $\sim9'$
and a width of $\sim1.5'$ in its central portion. The pulsar emission
is enclosed in a single pixel. The same is true for the brightest point
source in the field (marked as ``A''), an AGN which allowed us to estimate
the overall Galactic absorption (see text). Two
point sources are seen close to the southern end of the tail (marked as 
``1'' and ``2''). Multiwavelength data suggest they are unrelated
extragalactic objects.
\label{trail}}
\end{figure}

\clearpage

\begin{figure}
%\epsscale{.80}
\plotone{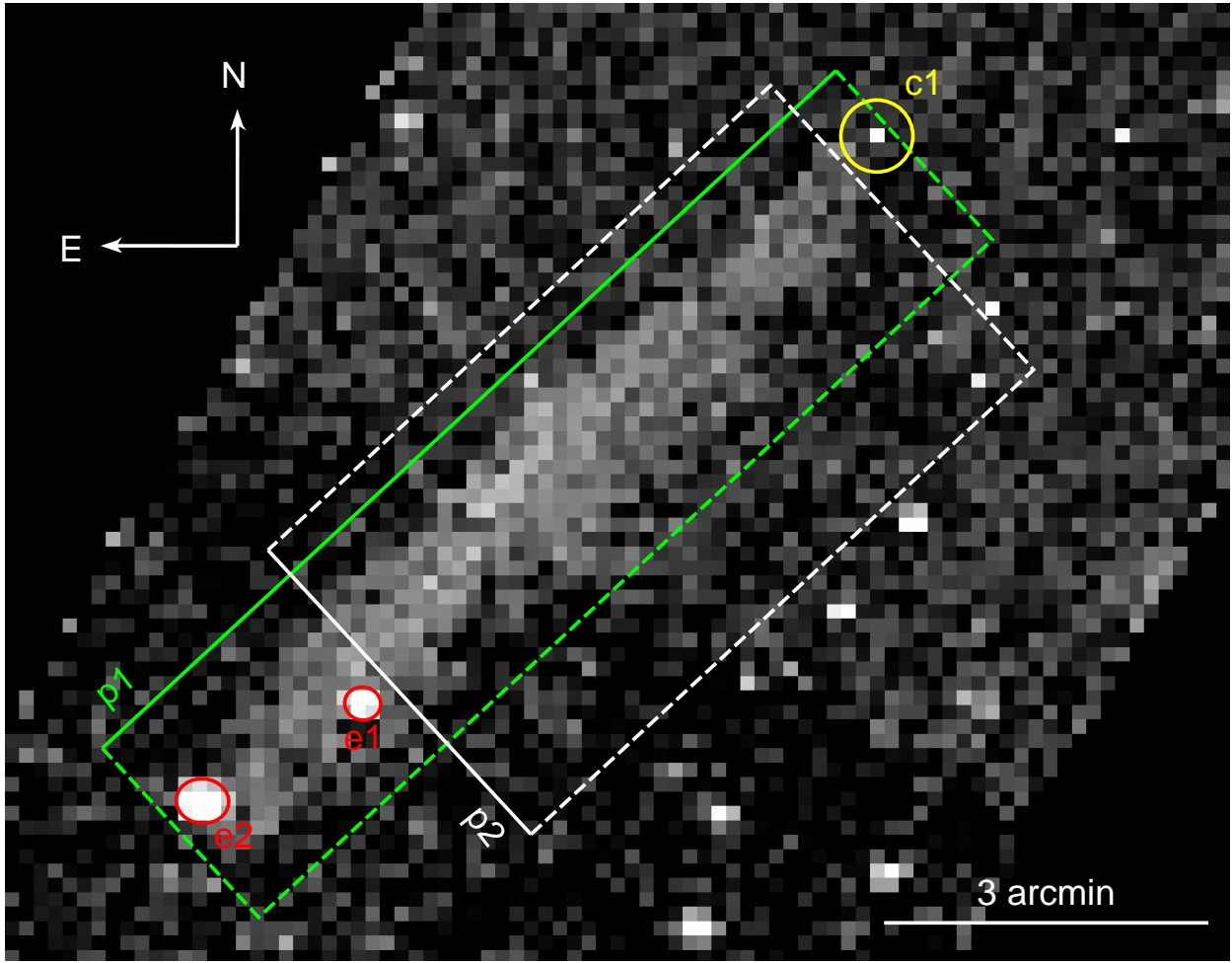}
\caption{Same as Fig.\ref{trail}. Regions used to generate surface 
brightness profiles for PSR J0357+3205 and its tail.
Circle c1 marks the 20$''$ radius region
from which we extracted
the radial profile of the pulsar counterpart shown in Fig.~\ref{psf}.
The regions from which
the brightness profiles of the tail (shown in Fig.~\ref{length} and Fig.~\ref{width}) were extracted are 
marked as p1 (along the main axis) and p2 (in the orthogonal direction).
Ellipses e1 and e2, computed using the {\em wavdetect} tool, were excluded
from the analysis to remove the counts from the
point-like sources ``1'' and ``2'' (see Fig.~\ref{trail}).  
\label{regions}}
\end{figure}

\clearpage

\begin{figure}
\epsscale{.60}
\plotone{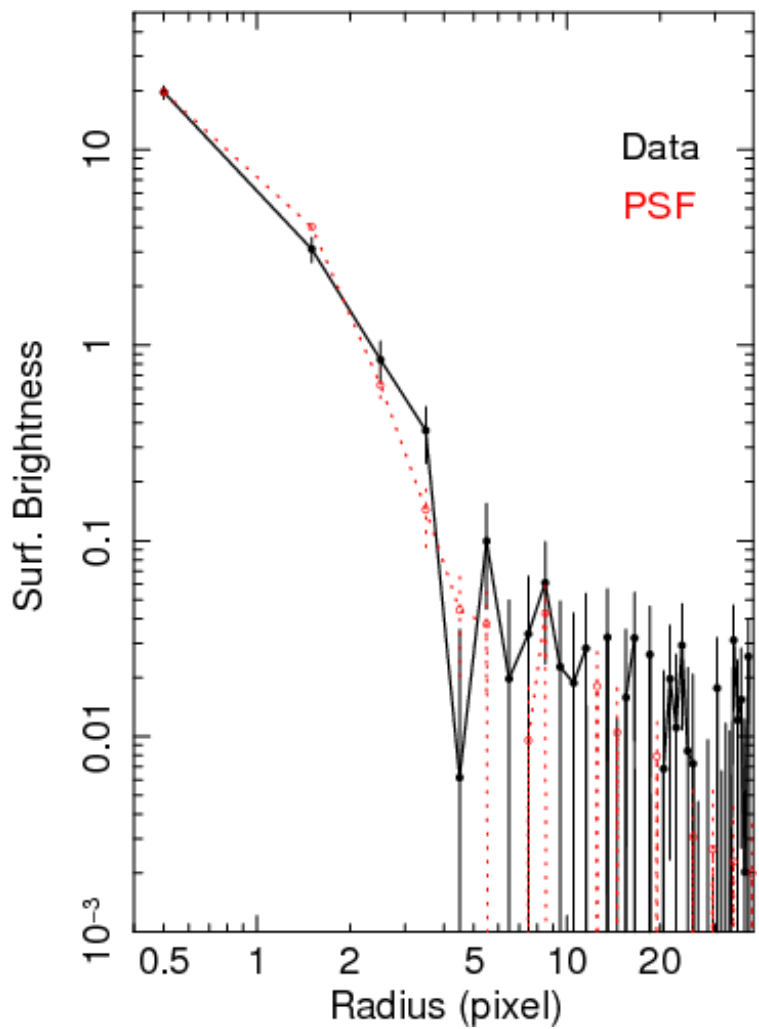}
\caption{Radial profiles (0.5-6 keV energy range) for the X-ray counterpart of PSR J0357+3205 
(background-subtracted, black points) and for a simulated point source 
(red points) having flux, spectrum
and detector coordinates coincident with the ones of the pulsar counterpart
(see text for details). The two profiles agree very well and there is no
evidence for significant diffuse emission in the surroundings of the pulsar
up to $20''$. 
\label{psf}}
\end{figure}

\clearpage

\begin{figure}
\epsscale{1.0}
\plotone{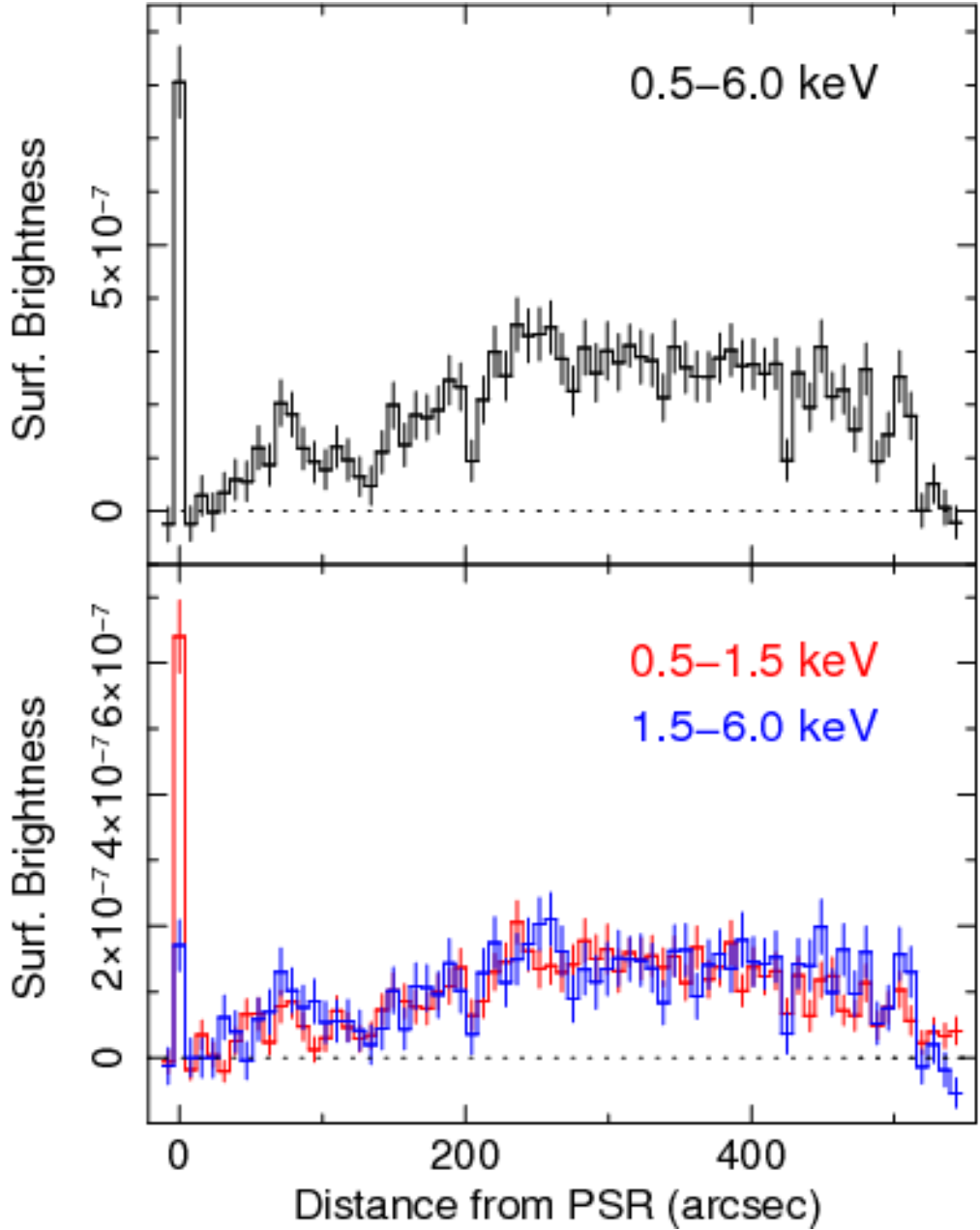}
\caption{Exposure-corrected, background subtracted surface
  brightness profiles of the tail along its 
main (North-West to South-East) axis (see also Fig.~\ref{regions}).  The upper panel 
shows the 0.5-6.0 keV energy range; the lower panel 
shows the 0.5-1.5 keV and 1.5-6.0 keV energy ranges.
The peak corresponding to PSR J0357+3205 is easily seen. Source ``1'' and ``2''
(see Fig.\ref{trail}) have been removed.
The rather smooth profile of the tail
as well as its broad maximum $\sim4.5'$ away from the pulsar is apparent. 
A possible local minimum in the
surface brightness is also seen at $\sim2'$ from the pulsar position.
The profiles in the soft (0.5-15.5 keV) and hard (1.5-6 keV) energy range
are almost indistinguishable.
\label{length}}
\end{figure}

\clearpage

\begin{figure}
\epsscale{1.0}
\plotone{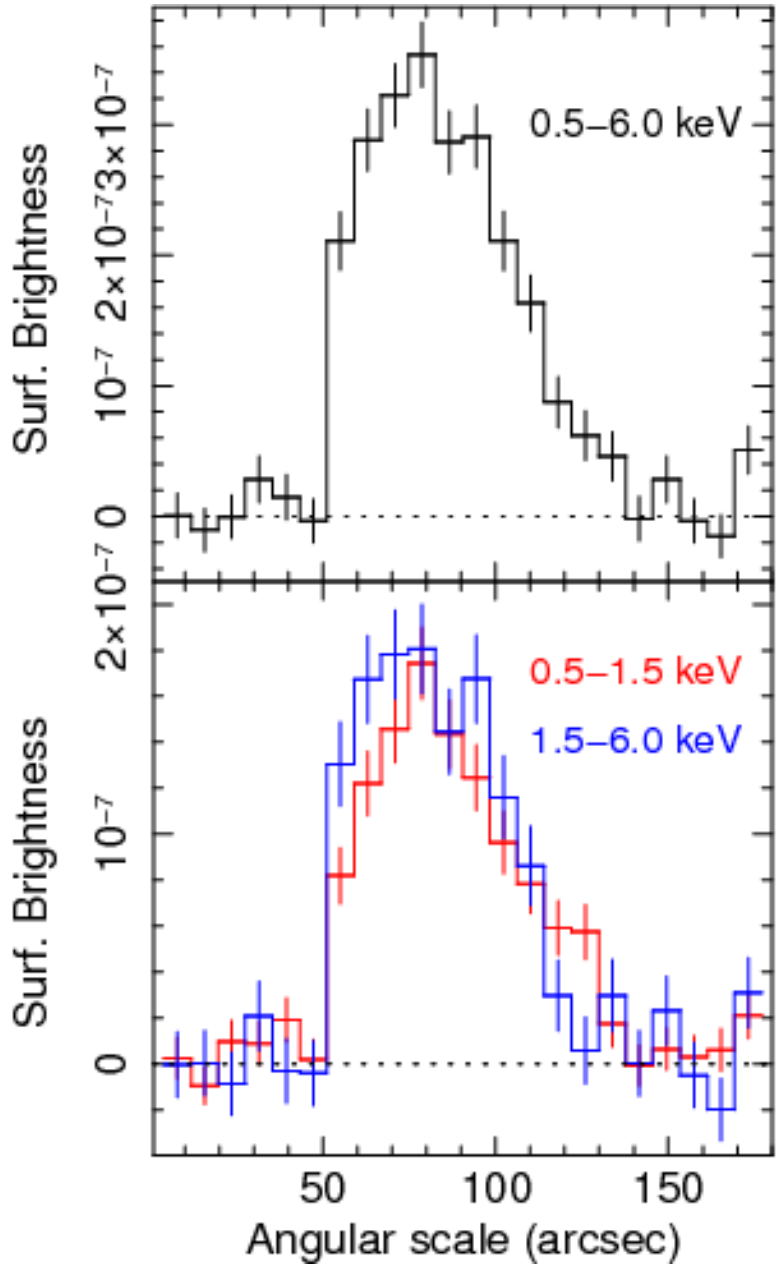}
\caption{Exposure-corrected, background subtracted surface
  brightness profiles of the tail along its width. 
%North-East is left (see also Fig.~\ref{regions}).  
The angular scale refers to the North-East to South-West direction
marked as ``p2'' in Fig.~\ref{regions}.
The sharp edge on the Northeastern
side is apparent, as well as the shallower decay on the opposite side. The
profiles in the soft (0.5-1.5 keV) and hard (1.5-6 keV) energy ranges
are very similar, with a slightly sharper edge in the hard band.
\label{width}}
\end{figure}

\clearpage

\begin{figure}
\epsscale{1.0}
\plotone{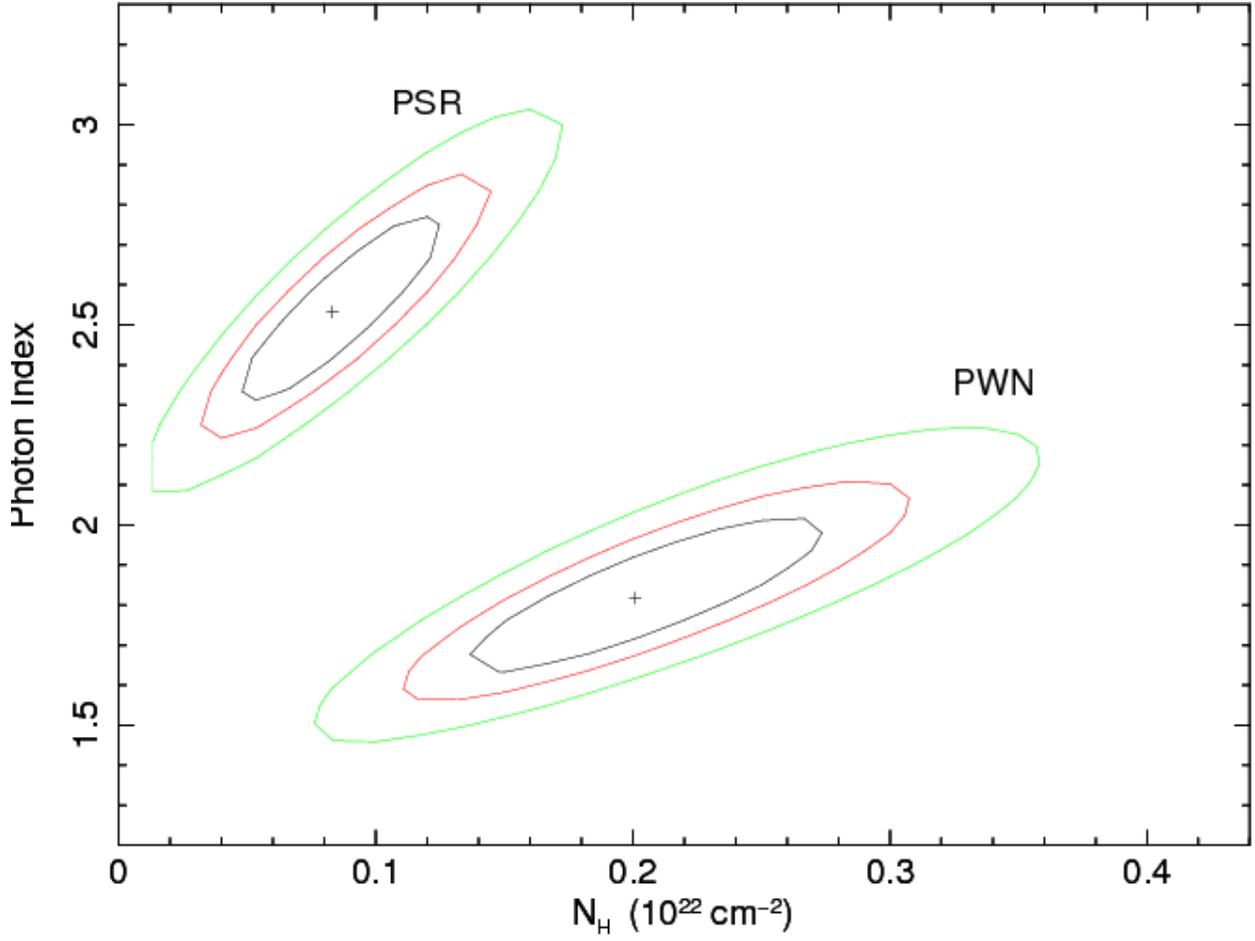}
\caption{Results of spectroscopy on the pulsar counterpart
as well as on the diffuse emission feature. Using an absorbed
power law model (see text), error ellipses (at 68\%, 90\% and 99\%
confidence level) for the
absorbing column $N_H$ and the photon index $\Gamma$ are shown.
\label{contours}}
\end{figure}

\end{document}